# Ejection of iron-bearing giant-impact fragments and the dynamical and geochemical influence of the fragment re-accretion

*Hidenori Genda[1,2], Tsuyoshi Iizuka[2], Takanori Sasaki[3,4],

Yuichiro Ueno[1,4], Masahiro Ikoma[2,5]

[1]*Earth-Life Science Institute, Tokyo Institute of Technology,*
*2-12-1 Ookayama, Meguro-ku, Tokyo 152-8551, Japan.*
[2]*Department of Earth and Planetary Science, The University of Tokyo,*
*7-3-1 Hongo, Bunkyo-ku, Tokyo 113-0033, Japan.*
[3]*Department of Astronomy, Kyoto University,*
*Kitashirakawa-Oiwake-cho, Sakyo-ku, Kyoto 606-8502, Japan.*
[4]*Department of Earth and Planetary Sciences, Tokyo Institute of Technology,*
*2-12-1 Ookayama, Meguro-ku, Tokyo 152-8551, Japan.*
[5]*Research Center for the Early Universe, The University of Tokyo,*
*7-3-1 Hongo, Bunkyo-ku, Tokyo 113-0033, Japan.*

*Corresponding author

Hidenori Genda

Earth-Life Science Institute, Tokyo Institute of Technology,

2-12-1-IE-14 Ookayama, Meguro-ku, Tokyo 152-8551, Japan.

Email: genda@elsi.jp

**Abstract:** The Earth was born in violence. Many giant collisions of protoplanets are thought to have occurred during the terrestrial planet formation. Here we investigated the giant impact stage by using a hybrid code that consistently deals with the orbital evolution of protoplanets around the Sun and the details of processes during giant impacts between two protoplanets. A significant amount of materials (up to several tens of percent of the total mass of the protoplanets) is ejected by giant impacts. We call these ejected fragments the giant-impact fragments (GIFs). In some of the erosive hit-and-run and high-velocity collisions, metallic iron is also ejected, which comes from the colliding protoplanets' cores. From ten numerical simulations for the giant impact stage, we found that the mass fraction of metallic iron in GIFs ranges from ~ 1wt% to ~ 25wt%. We also discussed the effects of the GIFs on the dynamical and geochemical characteristics of formed terrestrial planets. We found that the GIFs have the potential to solve the following dynamical and geochemical conflicts: (1) The Earth, currently in a near circular orbit, is likely to have had a highly eccentric orbit during the giant impact stage. The GIFs are large enough in total mass to lower the eccentricity of the Earth to its current value via their dynamical friction. (2) The concentrations of highly siderophile elements (HSEs) in the Earth's mantle are greater than what was predicted experimentally. Re-accretion of the iron-bearing GIFs onto the Earth can contribute to the excess of HSEs. In addition, the estimated amount of iron-bearing GIFs provides significant reducing agent that could transform primitive $CO_2$-$H_2O$ atmosphere and ocean into more reducing $H_2$-bearing atmosphere. Thus, GIFs are important for the origin of Earth's life and its early evolution.

# 1. Introduction

According to the classical model for planet formation called "the planetesimal accretion model", planets are formed in situ in a protoplanetary disk through successive accretion of km-sized planetesimals (Safronov, 1969; Hayashi et al., 1985). In the terrestrial planet region in our solar system, if the minimum-mass solar nebula (Hayashi, 1981) is assumed, several tens of Mars-sized protoplanets form (Wetherill, 1985; Lissauer, 1987; Kokubo and Ida, 1998) and they collide with each other to form Earth-sized planets (Chambers and Wetherill, 1998; Agnor et al., 1999). Collisions among protoplanets are called giant impacts, and this stage lasts for ~ 100 Myr (Kokubo et al., 2006; Jacobson et al., 2014). Especially, the last giant impact onto Earth likely produced the Moon, which meets geophysical and geochemical observations of Moon (Stevenson and Halliday, 2014). Although the detailed accretion process of protoplanets is a matter debate (Hayashi et al., 1985; Walsh et al., 2011; Lambrechts and Johansen, 2012; Levison et al., 2015), giant impacts seem to be a natural consequence for terrestrial planet formation.

Giant impacts are not simply merging events but also include complex events such as hit-and-run collisions (Agnor and Asphaug, 2004; Genda et al., 2012) and erosive collisions (Leinhardt and Stewart, 2012). These collisions also more or less produce ejected materials that are gravitationally unbound by protoplanets and orbit around a star. Using the impact conditions obtained by $N$-body orbital calculations of protoplanets (Kokubo and Genda, 2010), Genda et al. (2015a) estimated that a large amount of materials (several Mars masses in total) is ejected by giant impacts throughout the giant impact stage, which is consistent with the previous estimate (Stewart and Leinhardt, 2012). Ejection of materials by collisions seems to be ubiquitous not only during the giant impact stage, but also during the protoplanet formation stage (Bonsor et al., 2015; Carter et al., 2015).

If giant impacts take place in extrasolar systems, they would leave observational signatures. Indeed, detection of tens of infrared excess of solar-type stars with ages of 10–100 Myr has been reported recently (Zuckerman et al., 2011). This infrared excess is attributed to the infrared emission of dusts in a debris disk. These dusts are heated by the star (e.g., Oloffson et al., 2012), and these warm debris disks (> ~ 200 K) are located roughly 1 to several AU from the central stars, which corresponds to our terrestrial planet region. Thus, observed infrared excesses can be explained by debris produced during giant impact events (Jackson and Wyatt, 2012; Genda et al., 2015a; Kenyon et al., 2016).

Here, we focus on the ejected materials produced by giant impact events during terrestrial planet formation (Fig. 1); we hereafter refer to the ejected materials as giant-impact fragments (GIFs). The GIFs may have effects on the characteristics of formed terrestrial planets, which have not been well investigated. For example, while the current Earth's orbit is nearly circular, the proto-Earth likely had a highly eccentric orbit just after the end of the giant impact stage (Kokubo et al., 2006). We propose that the GIFs can decrease its eccentricity via dynamical friction with planets, provided a sufficient amount of GIFs are ejected and distributed in the terrestrial planet region. Moreover, if the GIFs contain metallic iron that is ejected from the cores of colliding differentiated protoplanets, re-accretion of the GIFs on the Earth should have effects on the redox state of the early Earth's environment because metallic iron acts as a reductant. Since almost all highly siderophile elements (HSEs; Re, Au, Os, Ir, Ru, Pt, Rh and Pd) are partitioned into the iron cores of protoplanets, re-accretion of metallic iron in GIFs would be a potential source of excess HSEs in the current Earth's mantle (Kimura et al., 1974; Chou, 1978).

Here we perform the self-consistent simulations of successive giant collisions among protoplanets and calculate the mass of GIFs with a focus on their iron contents. To that end, we have developed a hybrid code that simulates both the orbital evolution and collision process of protoplanets in the giant impact stage, based on the proto-type of a hybrid code presented in Genda et al. (2011). The details of the code are described in Section 2, and the simulation results are shown in Section 3. We show that the production and re-accretion of GIFs are inevitable during the giant impact stage. In Section 4, we discuss the dynamical effects of the GIFs on the Earth, and quantitatively examine the geochemical effects of re-accretion of GIFs on the Earth.

## 2. Hybrid Code for Giant Impact Stage

Our hybrid code consists of two parts: an *N*-body code for describing the orbital evolution of protoplanets around the Sun and an impact code for describing a collision between two protoplanets. In the *N*-body code, we use the modified fourth-order Hermite scheme with an individual time step (Kokubo and Makino, 2004). This allows us to calculate the orbits of protoplanets accurately, considering their mutual gravitational interactions without missing collisions between them. In the impact code, we utilize the code developed in Genda et al. (2015b). In this code, the standard smoothed particle hydrodynamics (SPH) method (Lucy, 1997) is used, which is a flexible Lagrangian method of solving hydrodynamic equations and has been widely used to simulate giant impacts between protoplanets (Canup, 2004).

The numerical simulation is started with the *N*-body code under a certain initial condition where protoplanets are located in the terrestrial planet forming region, and the orbital evolution of protoplanets is integrated. When the surfaces of any two protoplanets contact or overlap in the *N*-body code, we switch from the *N*-body code to the SPH code. By using impact parameters (the impact velocity and the impact angle) obtained by the orbital calculation and planetary conditions of the colliding protoplanets (their masses, core-mantle ratios, and spin states), we calculate the collision between two protoplanets. We use 10,000 SPH particles with equal mass for the two colliding protoplanets. During the impact simulation, the gravities of the Sun and the other protoplanets that act on the SPH particles are included, and orbital evolutions of the other protoplanets are also calculated.

Next, we analyze the outcome of the collision, and determine the masses of the gravitationally bounded clumps after the collision (Genda et al., 2015b), whose masses are defined as $M_1$, $M_2$, $M_3$ ... in order from the largest body. We set the lower limit of the number of SPH particles (= 10 particles) for a clump. We classify the impact event into two types by comparing the largest and second largest bodies ($M_1$ and $M_2$). If $M_1/M_2$ is less than 100, we define this collision as a hit-and-run impact, and we include both the largest and the second largest bodies as protoplanets in the subsequent *N*-body calculation. Otherwise, we define it as a single-main-body-remaining impact, and we only include the largest body as a protoplanet. It is noted that not only a simple merging impact, but also an impact that destroys the impactor (smaller protoplanet) are classified into a single-main-body-remaining impact.

We also determine the mass of GIFs ($M_{\mathrm{GIFs}}$) and its composition (i.e., iron fraction). Here we define the materials other than protoplanet(s) after the collision as GIFs. Therefore, $M_{\mathrm{GIFs}} = M_{\mathrm{tot}} - M_1$ for a single-main-body-remaining impact, and $M_{\mathrm{GIFs}} = M_{\mathrm{tot}} - M_1 - M_2$, where $M_{\mathrm{tot}}$ is the sum of mass for colliding two protoplanets. Some of the ejected fragments soon re-accrete onto the protoplanets, or orbit near the protoplanet(s), making a disk structure, but we do not regard these types of fragments as GIFs. By using the data of the mass, position, and velocity for post-impact protoplanets(s), we switch the code from the SPH code to the *N*-body code. For simplicity, we remove the ejected materials (i.e., GIFs) produced by each collision from the subsequent *N*-body calculation. We calculate the orbits of the protoplanets until the next collision. These iterative calculations are continued for 200 Myrs. In all the simulations below, we assume that no nebular gas remains. In this hybrid code, we can trace the changes in the mass, spin state (obliquity and spin rate), and composition (core-mantle ratio) of protoplanets during the giant impact stage.

Although introducing GIFs in *N*-body calculation would be important for many aspects, it is very CPU time consuming to carry out the *N*-body calculation with large number of fragments. As a first step, here we carried out the *N*-body simulations without GIFs, but we estimated the mass and composition of GIFs produced by each giant impact and we qualitatively discuss the effects of GIFs on the dynamical and geochemical aspects in Section 4.

For the initial condition of protoplanets, we follow the classical planetesimal accretion model. In the stage of protoplanet formation through successive accretions of planetesimals, the growth mode of a protoplanet is oligarchic, and its mass is given by the isolation mass (Kokubo and Ida, 2000). If the minimum-mass solar nebula (Hayashi, 1981) is assumed, several tens of Mars-sized protoplanets form in the terrestrial planet formation region. In this study, we put 16 Mars-sized protoplanets with a total mass of 2.3 $M_\oplus$ from 0.5 to 1.5 AU for the initial conditions, where $M_\oplus$ is the Earth mass. This initial condition is the same one that Kokubo and Genda (2010) used.

All the protoplanets are assumed to be initially differentiated with a 30wt% iron core and 70wt% silicate mantle. In our SPH simulations, we use the Tillotson EOS (Tillotson, 1962) with the parameter sets of granite for the mantle and of iron for the core (Melosh, 1989). Although peridotite is more appropriate for the mantle materials, there is no available parameter set of peridotite for the Tillotson EOS. Instead, for many giant impact simulations (e.g., Benz et al., 1987; Canup and Asphaug, 2001; Agnor and Asphaug, 2004), the parameter set of granite or basalt was used. We expect there to be no significant difference in behavior among the shocked states of peridotite, granite and basalt during a giant impact because they have similar density-pressure relations in shocked states (e.g., Artemieva and Ivanov, 2004).

## 3. Numerical Results of the Giant Impact Stage

We have performed ten runs with different initial angular distributions of the protoplanets. Figure 2 shows the orbital evolution of the protoplanets for 200 Myrs in the typical two runs (Runs 3 and 5). Considering the energy loss due to the removal of GIFs in our simulation, we have confirmed that the error of the total energy for the orbital calculation in 200 Myr is typically within 0.01%.

As shown below, Runs 3 and 5 are the typical cases resulting in a high and low content of metallic iron in GIFs, respectively. The number of protoplanets decreases with time, which indicates that a single-main-body remaining giant impact events happen. In these simulations for Runs 3 and 5, giant impacts occur 23 and 25 times, respectively. In total for 10 runs, 245 giant impacts occur. We found 52% of giant impact events (= 127/245) are hit-and-run events, which is consistent with the previous studies (49% in Kokubo and Genda, 2010, and 38% in Chambers, 2013).

The numbers of the planets that survived after 200 Myrs are 5 and 4 for Runs 3 and 5, respectively. From ten runs, we found that the average number and the standard deviation of the final planets is 4.2 ± 0.8, which is identical within the uncertainty to the previous results obtained by *N*-body calculations, that is, 3.4 ± 0.6 and 3.6 ± 0.8 reported in Kokubo et al. (2006) and Kokubo and Genda (2010), respectively. In detail, the average number obtained here is slightly larger than those in the previous studies. In our hybrid code, since we removed the ejected materials produced by each collision, the total mass of the protoplanetary system decreases with time. It has already been shown in Kokubo et al. (2006) that a lighter protoplanetary disk results in a larger number of final planets, which would be responsible for the larger average number obtained here. Our result (4.2 ± 0.8) is also consistent with the recent results obtained by *N*-body calculations with fragmentation model (4.2 ± 0.9 in Chambers (2013) and 3.8 ± 0.9 in Quintana et al. (2016)).

Figure 3 shows the mass and semi-major axis for all planets finally formed. Typically, 1 or 2 Earth-sized planets are formed in the middle of the terrestrial planet region (~ 1 AU), and another 2 or 3 small planets are formed at the both edges of the terrestrial planet region, which is consistent with the previous studies (Kokubo and Genda, 2010; Chambers, 2013). These small planets are likely survivors, that is, they experienced no giant impact or a few hit-and-run giant impacts.

Figure 4 shows the snapshots for three types of giant impacts happened in Run 3.

Those giant impacts correspond to the 8th (Coll#8), 19th (Coll#19), and 23rd collisions (Coll#23), respectively. Figure 5 shows the mass (and the number of SPH particles) of the largest and the second largest bodies ($M_1$ and $M_2$) and GIFs obtained by analyzing the collision outcome. In Coll#8, two Mars-sized protoplanets collide at a low velocity ($v_{imp}$ = 1.04 $v_{esc}$) with a 47-degree impact angle, where $v_{imp}$ and $v_{esc}$ are the impact velocity and the two-body surface escape velocity, respectively, and a 0-degree impact angle corresponds to a head-on impact. Two colliding protoplanets merge together, and a small amount of GIFs (2.2 × $10^{22}$ kg) is ejected. All SPH particles in GIFs are gravitationally separated from each other (Figure 5). Since each iron core completely merges, the GIFs contain no iron particles. In Coll#19, two Mars-sized protoplanets collide at a high velocity ($v_{imp}$ = 2.6 $v_{esc}$) with a 14-degree impact angle, and they do not merge, which is classified as a hit-and-run impact. In Figure 5, the largest and second largest clumps survive the giant collision. Due to the high impact velocity in Coll#19, a lot of GIFs (3.7 ×$10^{23}$ kg) are ejected, and GIFs consist of small 4 clumps and a lot of the other single SPH particles. GIFs are ejected not only from their mantles, but also from their iron cores. In this collision, 4.9% of the mass in GIFs is iron material. In Coll#23, different sized protoplanets collide at a high velocity ($v_{imp}$ = 3.2 $v_{esc}$) with a 33-degree impact angle. The impactor (smaller protoplanet) is pulverized, and almost all of the impactor's materials from both mantle and core (1.3 × $10^{24}$ kg) are ejected. Since $M_2$ is much smaller than $M_1$, this collision is classified as a single-main-body-remaining impact. On the other hand, the target core remains intact.

The number of SPH particles ($10^4$ particles) used here for giant impact simulations would not be enough to precisely estimate the masses of the largest and the second largest bodies and GIFs. According to Genda et al. (2015b), higher-resolution impact simulation of planetesimals tends to result in more erosive and disruptive collision. Here we carried out additional high-resolution simulations ($10^5$ and $10^6$ SPH particles) for three types of giant impacts (Coll#8, Coll#19, and Coll#23). Table 1 lists the total mass and composition of GIFs for different numerical resolutions.

For simple merging event (Coll#8), the mass of GIFs ($M_{GIFs}$) increases by a factor of 2, but it is still a small fraction of $M_{tot}$. For hit-and-run event (Coll#19), both $M_{GIFs}$ and the fraction of metallic iron in GIFs ($f_{Fe}$) increase by 50%. For impactor-destroying event (Coll#23), neither $M_{GIFs}$ nor $f_{Fe}$ depends on the numerical resolutions. Although the dependence of $M_{GIFs}$ on the numerical resolution is different among various types of giant impacts, $M_{\text{GIFs}}$ tends to increase with the numerical resolution. In this sense, the results obtained by using $10^4$ SPH particles give us the lower estimate in the mass of GIFs.

Figure 6 shows a plot of the mass of GIFs ejected by each giant impact in Runs 3 and 5. Sometimes, catastrophic high-impact-velocity collisions, which produce GIFs greater than 0.03 $M_\oplus$, take place. Figure 6 also shows the composition of the GIFs, that is, the fractions of rock and iron materials. We found that some giant impacts eject significant amounts of metallic iron (red bars). Even in the case of a low-velocity impact such as the canonical Moon-forming impact, it was reported that a small amount of iron material was ejected from the impactor's core into a proto-lunar disk (e.g., Canup and Asphaug 2001). It was also reported that high velocity collisions among planetesimals and protoplanets during the stage of protoplanet formation, which is followed by the giant impact stage, produce a lot of debris, and some of them are metallic iron from differentiated planetesimals (Carter et al. 2015). Therefore, it is reasonable to expect a significant amount of ejected metallic iron for intermediate or high velocity giant impacts.

It has been reported that the standard SPH method used here has difficulty in picking up hydrodynamical instabilities (e.g., Agertz et al. 2007) such as Rayleigh-Taylor and Kelvin-Helmholtz instability. These instabilities create the mixed layer at the core-mantle boundary to some extent, and would enhance the amount of metallic iron in GIFs. In this sense, the amount of ejected metallic iron obtained here would be a lower estimate.

Figure 7 shows the total mass of GIFs during the giant impact stage in each run. A large amount of materials (from 0.3 to 0.8 $M_\oplus$) are ejected. On average, GIFs of 0.51 $M_\oplus$ are produced, which is consistent with the previous estimate (0.42 $M_\oplus$) in Genda et al. (2015a), where they performed giant impact simulations by using the impact conditions obtained separately by the *N*-body calculations for the orbital evolution of protoplanets done by Kokubo and Genda (2010). Although the average mass of ejected materials produced by a single giant-impact event is small (a few percent of the two colliding protoplanets), the cumulative amount of the ejected materials is not negligible relative to the total mass of the protoplanetary system (2.3 $M_\oplus$). For each run, GIFs contain iron materials (0.6–25wt% of GIFs), and in particular, catastrophic high-impact-velocity collisions between protoplanets in Runs 3, 6 and 9 produce large amounts of ejected iron materials.

## 4. Dynamical and Geochemical Effects of GIFs on the Earth

*4.1. Circularization of the Earth's Orbit via Dynamical Friction of GIFs*

In many orbital simulations of protoplanets in the giant impact stage (Chambers and Wetherill, 1998; Kokubo et al., 2006), the eccentricities of the terrestrial planets finally formed are rather high (~ 0.1) compared with the present ones of the Earth and Venus (~ 0.01). Such high eccentricities are a natural consequence of the giant impact stage; otherwise protoplanets in separate orbits never undergo orbital crossing. To date, two mechanisms for damping such high eccentricities have been proposed: the drag force of the remaining nebular gas (Kominami and Ida, 2002) and the dynamical friction due to the planetesimals that are yet to accrete onto the protoplanets during the oligarchic growth stage (O'Brien et al., 2006). In Section 3, we have found that a large quantity of GIFs are inevitably produced during the giant impact stage, which suggests that the GIFs may be able to damp the high eccentricities of fully-grown terrestrial planets like remaining planetesimals can do. We should emphasize that our scenario is self-sufficient, so that we do not require the remaining nebular gas nor planetesimals during the giant impact stage to explain the current low eccentricities of the terrestrial planets.

According to Schlichting et al. (2012), remaining planetesimals with a total mass of 0.01 $M_\oplus$ can damp the excited eccentricity of the Earth to the present level. In the limit of efficient damping of eccentricities of planetesimals due to their mutual collisions, they estimated a total mass of the remaining planetesimals that is required to damp the eccentricities of the planets by comparing the eccentricity damping timescale for the terrestrial planets through dynamical friction with remaining planetesimals ($t_{\text{damp}}$) to the accretion timescale of the remaining planetesimals ($t_{\text{acc}}$). The condition $t_{\text{acc}} > t_{\text{damp}}$ must be fulfilled for the eccentricities of the planets to be fully damped. Such a condition is derived as

$$\Sigma_{\text{pl}} \gtrsim \Sigma_{\text{tp}} \left(\frac{v}{v_{\text{esc}}}\right)^2, \qquad (1)$$

where $\Sigma_{\text{pl}}$ and $\Sigma_{\text{tp}}$ are the mass surface densities of remaining planetesimals and the terrestrial planets, respectively, and $v$ and $v_{\text{esc}}$ are the velocity dispersion and escape velocity of the terrestrial planets, respectively. Since $v/v_{\text{esc}}$ is ~ 0.1–0.3 during the giant impact stage in by *N*-body simulations (Chambers, 2001), $\Sigma_{pl}/\Sigma_{tp}$ is estimated to be 0.01–0.1. Therefore, Schlichting et al. (2012) concluded that remaining planetesimals with roughly 0.01 $M_\oplus$ can damp the Earth's eccentricity.

Although detailed simulations including gravitational interactions between protoplanets and remaining small bodies such as GIFs are needed to confirm the above conclusion, dynamical friction by such a large quantity of GIFs—0.3–0.8 $M_\oplus$ obtained in our simulations—is a promising mechanism for damping the eccentricities of the excited terrestrial planets. Note that not all of the GIFs contribute to damping the eccentricities of formed planets. Some of the GIFs produced early in the giant impact stage re-accrete on protoplanets before the Earth is completely formed (i.e., the last giant impact), and small GIFs (~ μm) produced via mutual collisions among GIFs can also be easily blown out by solar radiation pressure. The half-life of such a debris disk produced by giant impacts is estimated to be the order of 10 Myr in the cases of both re-accretion (Jackson and Wyatt, 2012; Bottke et al., 2015) and blowing-out (Genda et al., 2015a). Therefore, the large fraction of GIFs produced early in the giant impact stage cannot survive until the last giant impact (~ 100 Myr), but the GIFs produced in the late stages can survive and would contribute to damping the Earth's eccentricity.

*4.2. GIFs as a Potential Source of Excess HSEs in the Earth's mantle*

Highly siderophile elements (HSEs) are strongly depleted in the Earth's mantle relative to chondrites, but their relative abundances are nearly chondritic despite their significantly different silicate-metal partitioning coefficients (Mann et al., 2012). The observed HSE signature has led to the widely held hypothesis that the HSEs were almost completely stripped from the Earth's mantle through metal-silicate segregation as well as sulfide-silicate segregation and, after core formation, a small amount of materials enriched in HSEs was supplied to the Earth (Kimura et al., 1974; Chou, 1978; Rubie et al., 2016) via the late-accretion (or "late veneer", which has sometimes been used as the synonym with "late-accretion"). As the source for the late veneer, chondrites in the outer asteroid belt (Drake and Righter, 2002) and leftover planetesimals after the giant impact phase (Schlichting et al., 2012) have been proposed. The former is further considered as a potential source of the Earth's volatiles (e.g., Albarède, 2009).

Here, as an alternative, we argue that the excess abundances of HSEs in the Earth's mantle can be attributed to the re-accretion of GIFs. The protoplanets gravitationally capture the GIFs on a timescale of 10 Myr after the end of each giant impact (Genda et al., 2015a), whereas the terrestrial magma ocean largely solidifies in several Myr around 1 AU (Hamano et al., 2013). There is a debate about the single timing for completion of Earth's core formation, that is, there would be multiple stages of Earth's core formation. Since giant impacts are one of the major processes that lead to core formation, here we assumed that major phase of core formation ended by the stage of the GIFs' re-accretion after the last giant impact.

According to our simulations, the GIFs contain 0.6–25wt% of iron (Fig. 7). Consequently, a substantial amount of HSEs in metallic iron materials of GIFs with nearly chondritic relative proportions would be delivered to the mantle by GIF re-accretion. For example, in Runs 1, 5 and 8 (the cases for low concentration of iron), GIFs of ~0.4 $M_\oplus$ are produced and their iron contents are a few percent. If 10% of the GIFs re-accrete onto the Earth after the last giant impact, the relative abundances of HSEs compared to CI chondrites are as much as $3 \times 10^{-3}$, which is consistent with the abundance of HSEs in the Earth's mantle ($1 \times 10^{-3}$ to $4 \times 10^{-3}$; Walker, 2009).

If the metallic iron of re-accreting GIFs simply sinks into the Earth core, HSEs in Fe of GIFs cannot be left behind in the Earth's mantle. A significant fraction of HSEs delivered by the GIF-metals should reside in the silicate Earth if the metals oxidized during or after the re-accretion process. The oxidation of re-accreted metals would occur on the Earth's surface if the Earth had ocean on the surface, and would occur in the Earth's mantle because the Earth's mantle would have been oxidized during the main phase of Earth's accretion and core growth as indicated by moderately siderophile element systematics (Wade and Wood, 2005; Rubie et al., 2011). The mantle oxidation could be caused either by disproportionation of ferrous Fe to ferric and metal Fe during silicate perovskite crystallization followed by the metal segregation (Wade and Wood, 2005; Frost et al., 2008) or by a transition from volatile-depleted to volatile-enriched accreting materials (Rubie et al., 2011).

The oxidative reaction efficiently proceeds especially when the size of metallic iron in GIFs is small. If the size of metallic iron in GIFs is planetary sized (~ 1000 km), most of re-accreting metallic iron cannot be oxidized. Due to the limit of the resolution for our numerical simulation, we cannot directly calculate the size of iron droplets ejected by giant impacts that is smaller than the size of a single SPH particle. Nevertheless, we can roughly estimate the typical size of iron droplets by considering the balance between surface tension of liquid iron and the local kinetic energy during fragmentation. According to Melosh and Vickery (1991), the typical size (*d*) of molten droplets ejected by an impact is given by

$$d \approx \left(\frac{40\sigma_{\text{surf}}}{\rho}\right)^{1/3} \left(\frac{D_{\text{imp}}}{v_{\text{imp}}}\right)^{2/3}, \qquad (2)$$

where $\sigma_{\text{surf}}$ is the surface tension (~ 2 N/m for liquid iron), $\rho$ is the density (~ 7000 kg/m$^3$), $D_{\text{imp}}$ is the impactor's diameter (~ 6400 km for a Mars-sized protoplanet), and $v_{\text{imp}}$ is the impact velocity (~ 10 km/s). We found $d \sim 17$ m, which is comparable to the original size of the iron meteorite that made Barringer Crater, Arizona, USA ($d \sim 40$ m). Most of parts of iron droplets with this size burn, and/or is pulverized into much smaller pieces during the Earth's atmospheric entry of this iron meteorite (Collins et al., 2005), or iron droplets are even partially vaporized at the impact onto ocean or sea floor (Kraus et al. 2015). Therefore, we expect that metallic Fe is efficiently oxidized by water on the Earth's surface and by ferric Fe in the mantle, allowing the late addition of HSEs to the Earth's mantle by GIFs.

The presence of oceans as far back as 4.3 Ga is indicated by heavy oxygen isotope ratios in Hadean zircons (Cavosie et al., 2005). Yet several recent lines of evidence suggest the existence of water even before the Moon-forming impact. For example, the discovery of water in lunar anorthosites indicates that the lunar magma ocean contained water up to 300 ppm (Hui et al., 2013). Moreover, the D/H ratios of lunar volcanics that are similar to that of Earth's seawater (Saal et al., 2013) imply a common origin for the water on the Earth and Moon. Noble gas abundance and isotopic compositions suggest that Earth acquired most of its major volatile elements by accretion of volatile-rich planetesimals and/or protoplanets (Halliday, 2013; Dauphas and Morbidelli, 2014). The presence of water in the planetesimals that eventually formed the Earth is also supported by recent theories for the thermal structure of protoplanetary disks (Oka et al., 2011; Min et al., 2011). Thus, the inference that the Earth had some amount of ocean during the stage of GIFs re-accretion is reasonable.

Since GIFs also accrete onto Mars and the Moon, HSEs would be supplied in their mantles. It is reasonably well accepted that the abundance of HSEs for the Martian mantle is roughly similar to that of the Earth's mantle, and that for the lunar mantle is about 20 times lower than that of the Earth's mantle (Walker, 2009; Dale et al., 2012), although there is a debate whether the HSE signatures can be naturally explained by their partitioning between the silicate and metal (Righter et al., 2015; Sharp et al. 2015) or require late accretion (Day et al., 2016).

Assuming that all HSEs are supplied via the late accretion, the ratios of accreted mass for Earth/Mars and Earth/Moon are estimated as 12–23 and 200–700, respectively (Schlichting et al., 2012). The gravitationally enhanced cross section of a planet $\sigma$ is given by

$$\sigma = \pi R^2 \left[1 + \left(\frac{v_{esc}}{v_{rel}}\right)^2\right], \quad (3)$$

provided $v_{rel}^{1/2} > \sim 2 \, (M_p/M_*)^{1/3}$ (Ida and Nakazawa, 1989), where $v_{rel}$ is the relative velocities of GIFs, $M_p$ and $M_*$ are the masses of the protoplanet and central star. If $v_{rel} \ll v_{esc}$, the ratio of the cross section between planet 1 and 2 is written as

$$\frac{\sigma_1}{\sigma_2} = \frac{\rho_1}{\rho_2}\left(\frac{R_1}{R_2}\right)^4. \quad (4)$$

When the radii and densities of the Earth and Moon are applied to this equation, the ratio of the cross sections is 300, which is consistent with the late accretion mass ratio inferred for the Earth and Moon. If this is also applied for the Earth and Mars, the ratio of the cross section is about 17, which is also consistent with the observed value.

Since the age of core formation on Mars (~ 2 Myr, Tang and Dauphas, 2014; Dauphas and Pourmond, 2011) is much earlier than the Moon-forming giant impact onto the Earth (> 30 Myr, Kleine et al., 2009), Mars would be a survivor of protoplanets and would have been exposed to more GIFs during the entire giant impact stage than the Earth and Moon. Meanwhile, Mars should be formed at the outer edge of the terrestrial planet region where giant impact events rarely or never happen. Therefore, total mass of GIFs at Mars' orbit should be much smaller than that at Earth's orbit. Although longer exposure time of GIFs for Mars enhances the accreted mass of GIFs, lower surface density of GIFs reduces the accreted mass.

As discussed above, the estimated ratios by late accretion with strong gravitational focusing can account for HSE contents in the mantle of the Moon as well as Mars, provided that the relative velocities between the terrestrial planets and accreted materials are significantly low compared to the escape velocities. Such low relative velocities can be achieved by small objects available in the terrestrial planet region rather than those delivered from the outer asteroid belt.

Furthermore, the late accreted small objects could damp the original high eccentricities of terrestrial planets (see Section 4.1). Note, however, that if the late veneer has a chondritic composition, accounting for the HSE abundances and eccentricities of the terrestrial planets at the same time requires a typical size of ≤10 m for the late accreted objects in order to ensure efficient damping of the relative velocities of planetesimals (Schlichting et al., 2012), which seems to be too small for leftover planetesimals after the giant impact stage. The GIF late accretion hypothesis can circumvent this problem. The GIFs would originate mainly from silicate portions of differentiated protoplanets, and therefore have lower HSE abundances than chondrites (Fig. 7). This relaxes the upper limits for the total mass and typical size of the late accreted objects.

*4.3. Reduction of the primordial atmosphere and ocean by re-accretion of GIFs*

The re-accretion of GIFs on the Earth after the solidification of magma ocean also has implications for the redox state of the Earth's earliest atmosphere. After core-mantle differentiation, oxygen fugacity of the Earth's uppermost mantle would have been defined by fayalite-magnetite-quartz (FMQ) buffer, in which $H_2O$ and $CO_2$ would dominate the magmatic volatile (Trail et al., 2011; Delano, 2001; Frost et al., 2008). The primordial $CO_2$-rich atmosphere and the ocean may have been reduced by the impact of considerable amount of GIFs, because the average composition of the metal-bearing GIFs should be more reducing than the FMQ mantle.

Under the heavy rain of the metal-bearing GIFs during the first 100 Myr, metal part of the GIF would have reacted with the ocean via the reaction, $Fe + H_2O \rightarrow FeO + H_2$. In reality, however, 75–99% of the GIFs are not metal but rocky fragments. Therefore, the reactions between GIFs and the primordial ocean may create more complex volatiles, which is beyond the scope of this paper. Instead, here we evaluate the reducing capacity provided by the re-accretion of GIFs by estimating the amount of $H_2$ produced by the impact of GIFs.

First, we assume that the average composition of the GIFs was similar to that of ordinary chondrites that is considered as the representative source materials of terrestrial planets before the core-mantle differentiation (Rubie et al., 2011). According to the equilibrium calculation of Schaefer and Fegley (2010), the impact of ordinary chondrite would create the highly reducing gasses ($H_2$ ~45%, CO ~25%, $H_2O$ ~20%, $CO_2$ ~5%) and other trace components including $N_2$ and $CH_4$. These volatiles could represent averaged gas composition created by the GIFs and ocean because our simulation suggests that the impact velocities of GIFs were high enough to vaporize themselves.

In this case, roughly 70% of the reacting water is converted into $H_2$ and remaining 30% stays in $H_2O$. Also, the total amount of $H_2$ by the GIFs' impact largely depends on the initial volume of the ocean before the impact, because the amount of GIFs was large enough to react with all of the pre-existing ocean as discussed in Section 4.2. As a result, provided that the initial mass of ocean was 3.5 times larger than present, 70 bar $H_2$ atmosphere would be produced during the GIF bombardment, and the present-day volume of ocean was left behind after the impact. This amount of $H_2$ could have sustained over ~ 200 Myr, when considering subsequent hydrogen escape into space (see Appendix).

A much higher $H_2$ pressure of atmosphere is possible depending on higher initial ocean mass as well as on higher metal fraction of GIFs. These parameters are highly uncertain, though the simple calculation suggests that very reducing atmosphere could have been sustained from 200 to possibly over 1,000 Myrs timescale after the cessation of GIF bombardment. Thus, when considering the re-entry of large amount of GIF, the Hadean and possibly Archean Earth would have highly reducing atmosphere despite of oxidized volcanic gas emission. This is compatible with the observed mass-independent isotope fractionation of sulfur isotopes in Archean sedimentary rocks, which requires a very low atmospheric-oxygen level and sufficiently high $CH_4$ or CO level (Pavlov and Kasting, 2002; Ueno et a., 2009).

## 5. Conclusions

In our solar system, many giant impacts among protoplanets occurred during terrestrial planet formation. These giant impacts more or less eject materials into space. In this study, we have simulated the stage of giant impacts by using a hybrid code that consistently deals with the orbital evolution of protoplanets around the Sun and the detailed process of a giant impact between two protoplanets. The results revealed that a significant amount of giant impact fragments (GIFs)—corresponding to ~ 0.5 $M_\oplus$ in total—are ejected by giant impacts throughout the giant impact stage. We found that GIFs contain a significant amount of metallic iron materials (0.6–25wt% of GIFs), which are ejected from the colliding protoplanets' cores.

Based on the numerical results, we have indicated that the GIFs have significant effects on the dynamical and geochemical characteristics of formed terrestrial planets: (1) Dynamical friction caused by GIFs left after the last giant impact can lower the eccentricity of the Earth from a highly eccentric orbit just after the last giant impact (its eccentricity is ~ 0.1) to its current value (~ 0.01). (2) Re-accretion of iron-bearing GIFs onto Earth can account for the excess of highly siderophile elements (HSEs) in the Earth's mantle as the source of the late veneer. (3) The estimated amount of iron-bearing GIFs provides significant reducing agent that could transform primitive $CO_2$-$H_2O$ atmosphere and ocean into more reducing $H_2$-bearing atmosphere. Early reducing atmosphere is important for the origin of Earth's life and its early evolution.

**Appendix. Calculation of lifetime of $H_2$ atmosphere against H-escape into space**

In order to evaluate the reducing capacity derived from the impact of GIF, we have calculated how long the $H_2$-atmosphere could have persisted against hydrogen escape into space. Since we consider a $H_2$-dominated atmosphere reduced by GIF's metallic iron in this paper, the assumption that the escape flux of $H_2$ is controlled by EUV energy flux (called energy-limited hydrodynamic escape) is valid (Watson et al., 1981):

$$\phi_{H_2} = \frac{\varepsilon f_{\mathrm{EUV}}(t) R}{4 G M_\oplus m_{H_2}} \quad [m^{-2}\, s^{-1}], \tag{A1}$$

where $\phi_{H_2}$ is the escape flux of $H_2$, $G$ is the gravitational constant, $R$ is the planetary radius, and $m_{H_2}$ is the mass of a hydrogen molecule. Escape efficiency $\varepsilon$ represents how large a proportion of received EUV energy is available for escaping hydrogen molecules. We adapt 0.3 for $\varepsilon$ (Sekiya et al., 1980). The EUV energy flux $f_{\mathrm{EUV}}(t)$ would have been several orders of magnitude greater than its present value (Zahnle and Walker, 1982) at the time of GIF re-entry. We used the following time-dependent model for the EUV energy flux (Ribas et al., 2005),

$$f_{\mathrm{EUV}}(t) = 0.03 \left(\frac{t}{10^9 \mathrm{yr}}\right)^{-1} \quad [\mathrm{W/m^2}]. \tag{A2}$$

We consider only $H_2$ for the atmospheric species, ignoring the other species for simplicity. By using Eqs. (A1) and (A2), we can estimate the timescale for $H_2$ atmospheric loss. If the partial pressure of $H_2$ in the atmosphere is 70 bar, it takes about 200 Myr for the $H_2$ atmosphere to be completely lost. The presence of CO and $CO_2$ in the upper atmosphere causes a decrease in the escape efficiency ($\varepsilon$) due to their efficient IR emission (Lammer et al. 2006), which extends the timescale for $H_2$ atmospheric loss. In this sense, the timescale estimated here (~200 Myr) would be a lower one, although the EUV energy flux is also quite uncertain.

**Acknowledgments**

We thank E. Asphaug and D. Stevenson for their helpful reviews, which led us to greatly improve this paper. We also thank anonymous referees for their constructive reviews. This research was supported by a grant for the Global COE Program entitled "From the Earth to "Earths"", MEXT, Japan. We thank the leader of the Global COE Program, Professor Shigeru Ida, for providing the opportunity for us to collaborate on this interdisciplinary work.

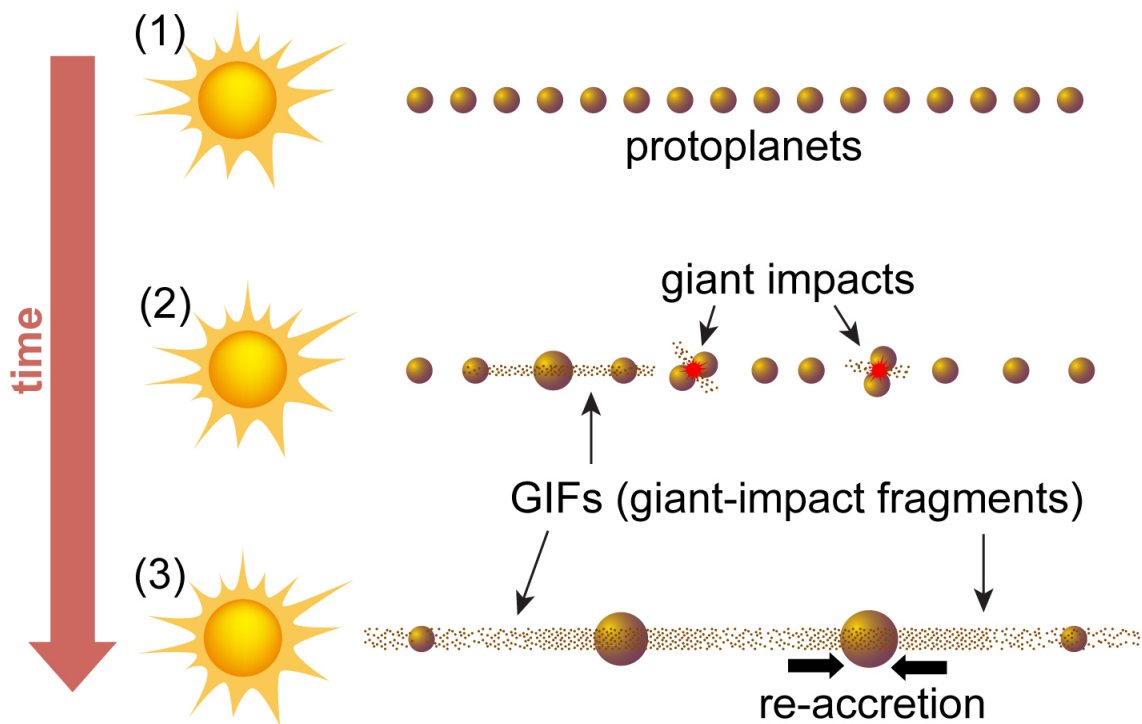

Figure 1. Schematic view of production and re-accretion of giant-impact fragments (GIFs). (1) Several tens of Mars-sized protoplanets form in the terrestrial-planet formation region. (2) These protoplanets collide with each other and eject large amounts of material into the orbital region. (3) The GIFs are re-accreted onto the terrestrial planets on a timescale of 10 Myr after the end of giant impacts. In this study, we simulate orbital evolutions and collisional processes of the protoplanets, and estimate the amount of GIFs and their compositions. We also discuss the dynamical and geochemical influence of the re-accretion of GIFs based on analytical arguments.

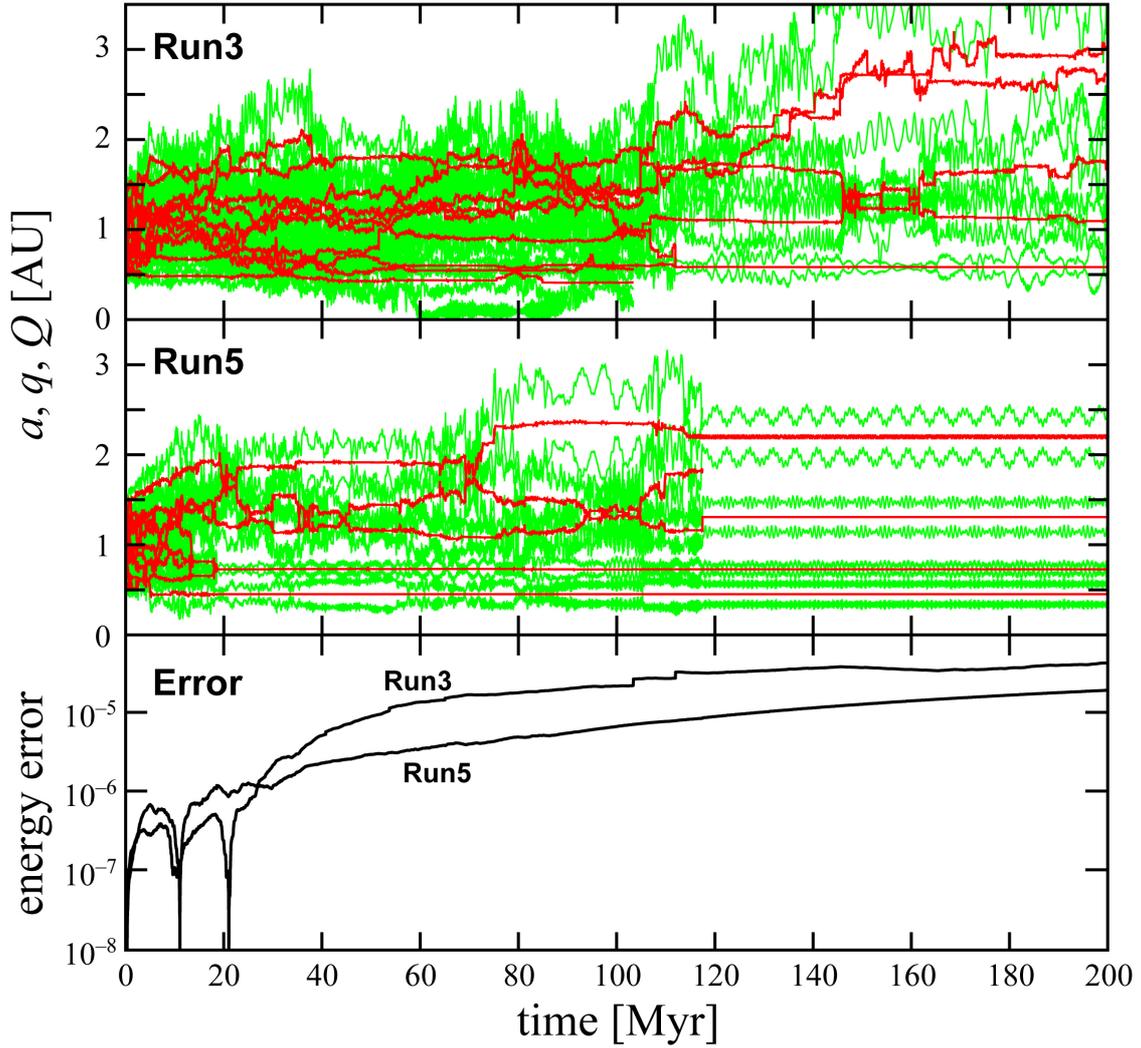

Figure 2. Orbital evolutions of protoplanets for two different initial angular distributions of protoplanets (Runs 3 and 5). Top and middle panels show temporal changes in the semimajor axes $a$ (red curves) and pericenter $q$ and apocenter $Q$ distances (green curves) of protoplanets for Runs 3 and 5, respectively. Bottom panel shows the error of the total energy (potential and orbital energies of protoplanets) for the orbital calculation, considering the energy loss due to the removal of the ejected materials in our simulation.

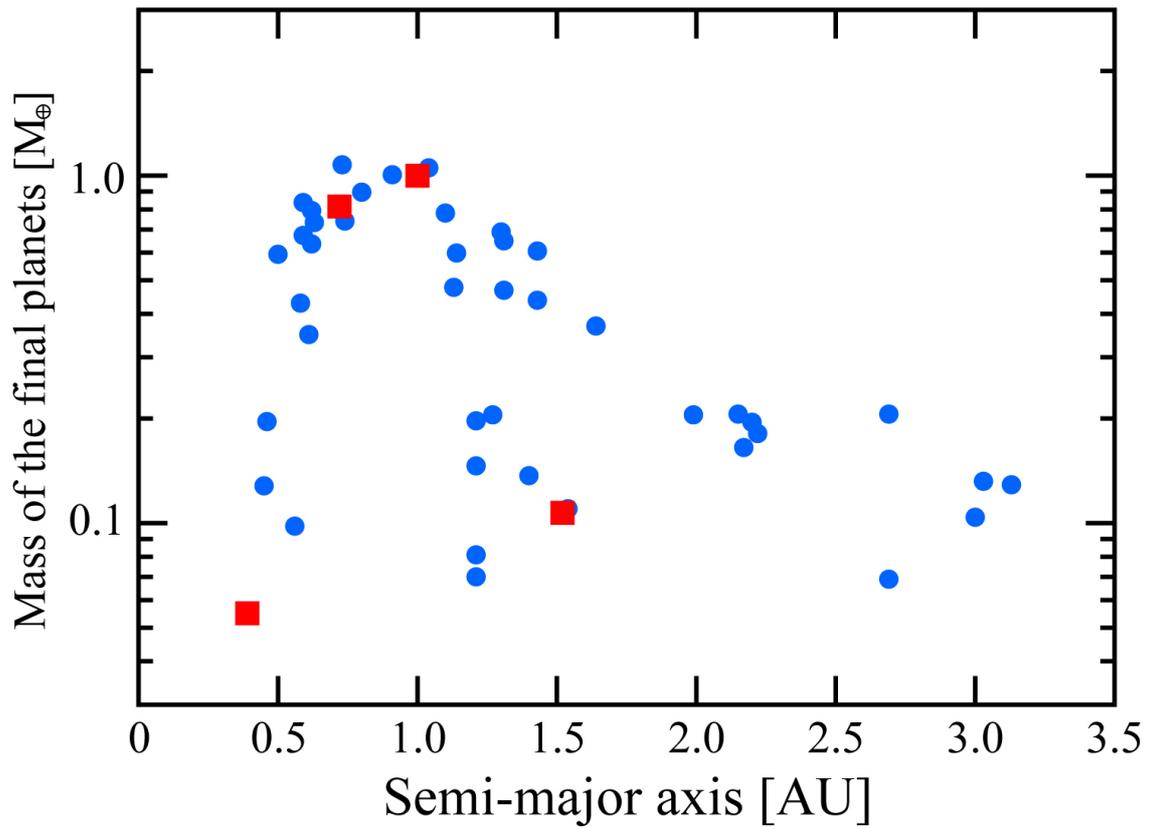

Figure 3. Mass versus semi-major axis for all protoplanets at the end of the simulations (*t* = 200 Myrs). Circles indicate numerical results, while square symbols show the terrestrial planets of our solar system for comparison.

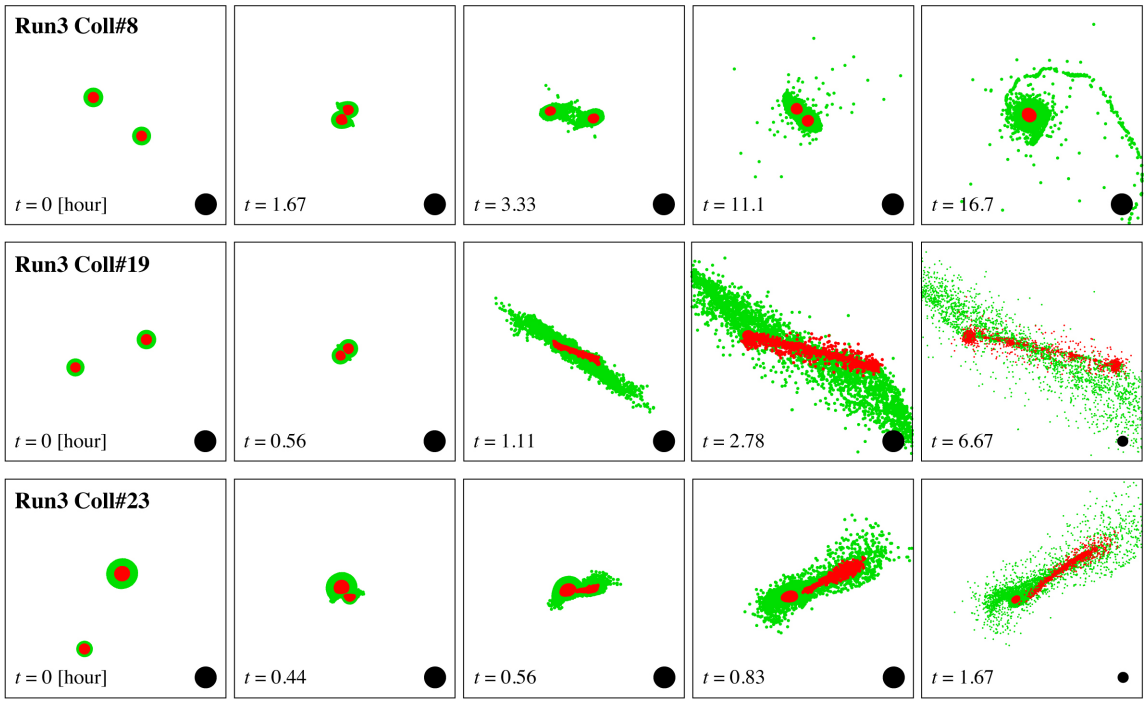

Figure 4. Snapshots of three giant impacts (Coll#8, #19, and #23) taking place in Run 3. Green and red particles represent mantle materials and iron materials, respectively. The filled black circles in the bottom-right corner of each panel are 5,000 km in diameter.

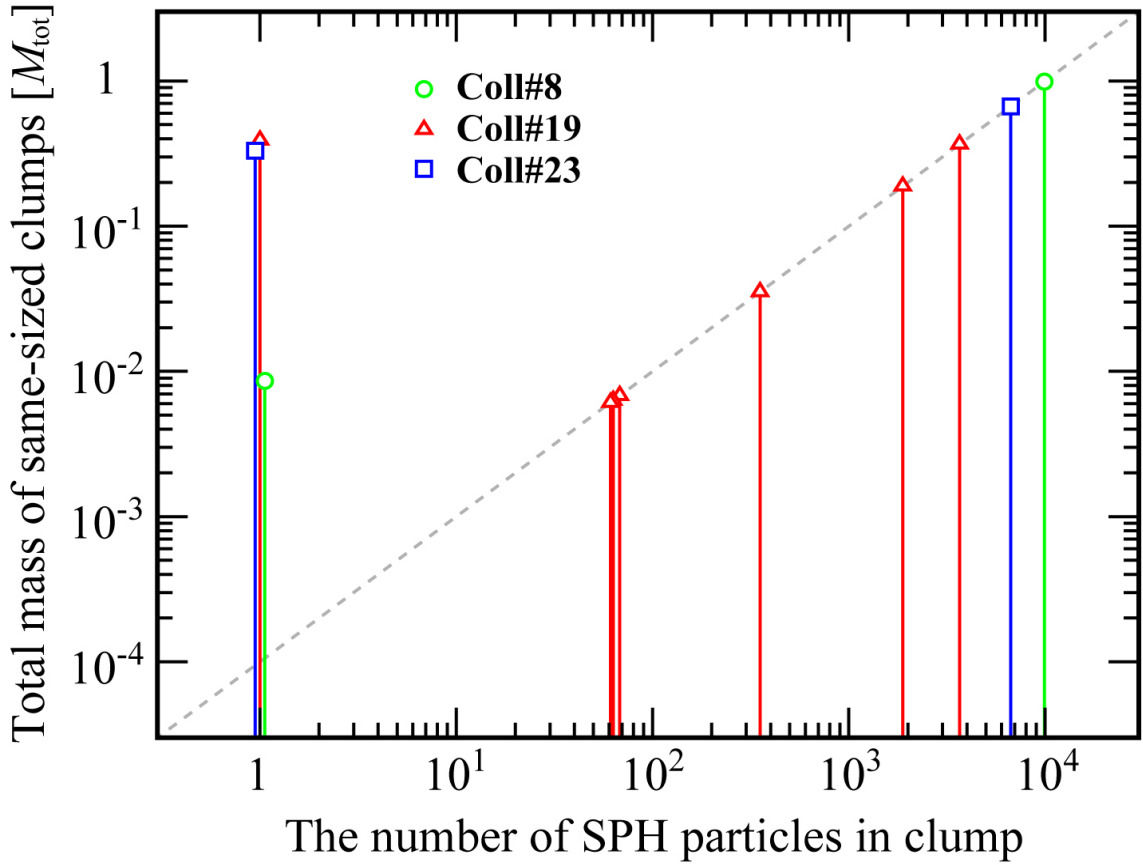

Figure 5. The mass of the largest and second largest bodies, and GIFs after giant impacts. Three giant impacts (Coll#8, #19, and #23) are considered. If the data points are on the dashed line, these clumps have different number of SPH particles. The data points that have 1 in the value of horizontal axis correspond to total mass of single SPH particle.

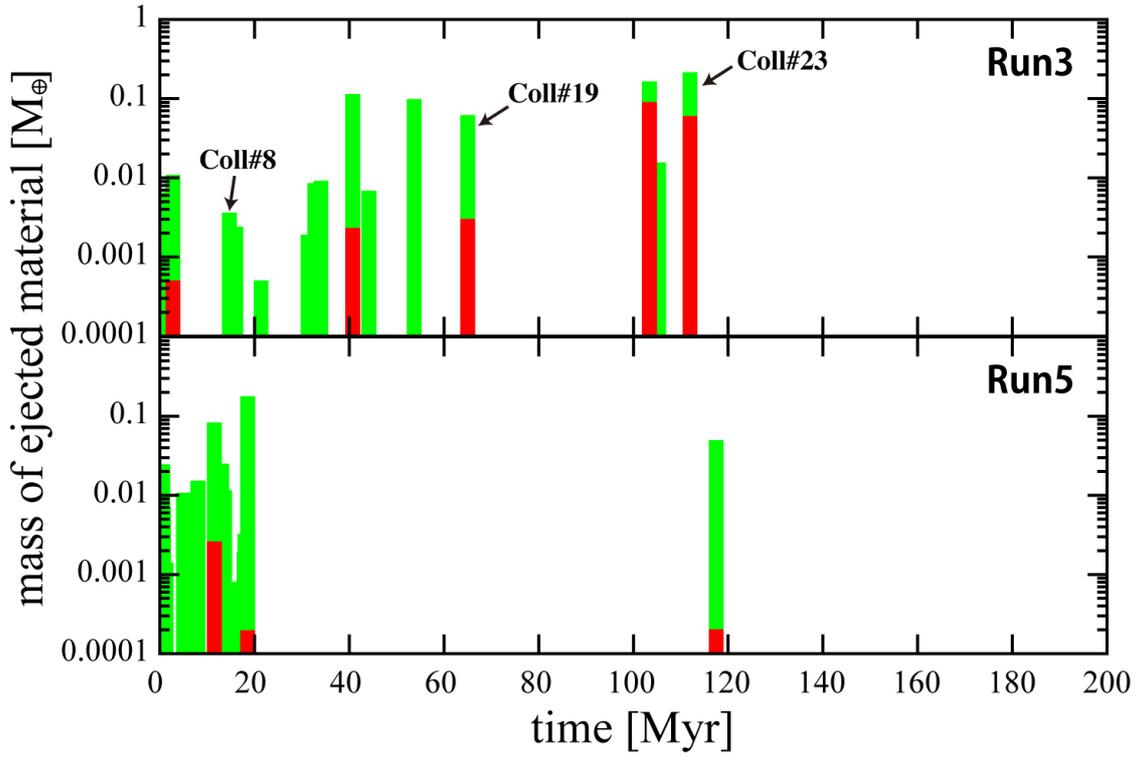

Figure 6. The total mass of ejected materials by each giant impact during 200 Myrs for Runs 3 and 5. The green bars represent the total mass of ejected materials, while the red bars represent that of iron components. Three bars indicated by Coll#8, #19 and #23 represent collisions shown in Fig. 4.

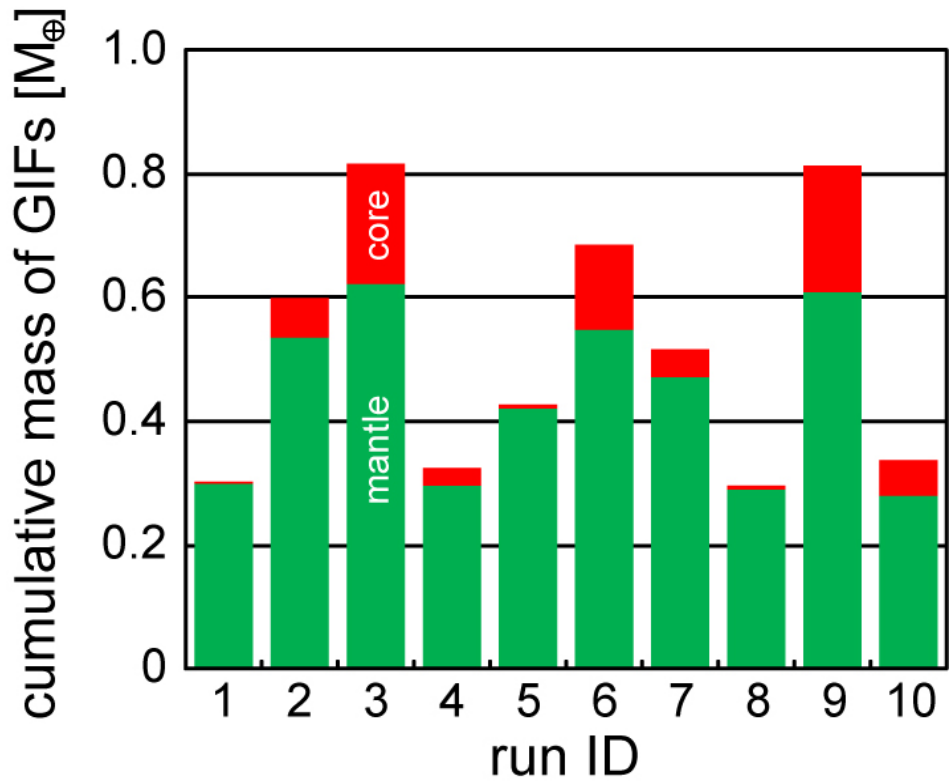

Figure 7. Cumulative mass of the GIFs in Earth mass units for ten runs with different initial orbital configurations of protoplanets. The colors of each bar show the composition of the ejected materials (green: rocky material from the protoplanets' mantles, red: iron material from their cores).

Table 1. Total mass of GIFs and its iron content for various numerical resolutions

| # of SPH particles | Coll#8 | | Coll#19 | | Coll#23 | |
|---|---|---|---|---|---|---|
| | $M_{GIFs}$ [$M_{tot}$] | $f_{Fe}$ | $M_{GIFs}$ [$M_{tot}$] | $f_{Fe}$ | $M_{GIFs}$ [$M_{tot}$] | $f_{Fe}$ |
| $10^4$ | 0.009 | 0% | 0.45 | 4.9% | 0.32 | 27.2% |
| $10^5$ | 0.012 | 0% | 0.52 | 6.9% | 0.33 | 27.1% |
| $10^6$ | 0.017 | 0% | 0.61 | 8.8% | 0.33 | 27.5% |

Note: $M_{GIFs}$ is the total mass of GIFs ejected by each giant impact, $M_{tot}$ is the summation of the target's and impactor's mass, and $f_{Fe}$ is the fraction of metallic iron in GIFs.